\documentclass[conference]{IEEEtran}
\IEEEoverridecommandlockouts

\usepackage{cite}
\usepackage{amsmath,amssymb,amsfonts,bm,mathtools}
\usepackage{graphicx}
\usepackage{booktabs,array,makecell,multirow}
\usepackage{xcolor}
\usepackage{siunitx}
\usepackage{textcomp}
\usepackage{microtype}
\usepackage{url}
\usepackage[hidelinks]{hyperref}
\usepackage{enumitem}
\usepackage[utf8]{inputenc} 

\def\BibTeX{{\rm B\kern-.05em{\sc i\kern-.025em b}\kern-.08em
    T\kern-.1667em\lower.7ex\hbox{E}\kern-.125emX}}

\hypersetup{
  colorlinks=true,
  linkcolor=blue!50!black,
  citecolor=blue!50!black,
  urlcolor=blue!50!black
}

\hypersetup{
  pdfauthor={Xiaolong Deng},
  pdftitle={Runtime Calibration as State-Trajectory Feedback Control in Quantum-Classical Workflows},
  pdfsubject={Runtime calibration for quantum-classical workflows}
}

\makeatletter
\newcommand{\linebreakand}{%
  \end{@IEEEauthorhalign}
  \hfill\mbox{}\par
  \mbox{}\hfill\begin{@IEEEauthorhalign}
}
\makeatother

\begin{document}

\title{Runtime Calibration as State-Trajectory Feedback Control in Quantum-Classical Workflows}

\author{\IEEEauthorblockN{Xiaolong Deng}
\IEEEauthorblockA{
\textit{Leibniz Supercomputing Centre}\\
Garching, Germany \\
xiaolong.deng@lrz.de
}
}

\maketitle

\begin{abstract}
In superconducting devices running variational workloads, gate and readout fidelities drift on hour timescales, while existing runtime schedulers treat backend quality as static. The temporal dimension of calibration remains unresolved. We formulate runtime calibration as a state-trajectory feedback-control problem under a fixed wall-clock budget, and investigate whether spending time on calibration now can improve the future optimization trajectory. Calibration quality proxy is represented as a drifting equivalent-age state, recovery action is modeled as costly state reset, and policies are evaluated by time-integrated optimization gap over the full execution window. Using a finite-horizon rollout controller, we compare feedback calibration against a strengthened family of open-loop baselines across three latency regimes: cloud-like (25 ms), local-millisecond (1 ms), and tight-loop (4 $\mathrm{\mu}$s). The results show a clear ordering: cloud-like feedback is generally uncompetitive, while local-ms and tight-loop regimes open a positive-gain region that grows with workload quality-sensitivity and initial calibration age. Crucially, the gap between local-ms and tight-loop control is modest for single-target recovery. The advantage of tight-loop integration emerges under capacity pressure, when many calibration targets must be processed within the same control window.
\end{abstract}


\section{Introduction}
\label{sec:intro}
Hybrid high-performance computing and quantum computing (HPC–QC) integration is rapidly transitioning from standalone prototypes to integrated accelerator-based architectures within leadership supercomputing centers, enabling complex variational workflows at scale~\cite{britt2017high,humble2021quantum,elsharkawy2023}. In such workflows many quantum circuits are executed inside classical control or optimization loops. Variational quantum algorithms (VQAs)~\cite{farhi2014,TILLY20221} are a representative example. Quantum executions are interleaved with classical preprocessing, parameter updates, and data movement, often exhibiting multi-scale characteristics on the wall-clock time of shared HPC--QC platforms. In superconducting devices~\cite{krantz2019,arute2019quantum}, qubit frequencies, readout responses, and gate fidelities drift on timescales ranging from minutes to hours due to environmental fluctuations and two-level-system dynamics~\cite{kandala2019,marciniak2026}. 

This drift leads to a runtime-system problem. As the workload progresses, degraded calibration can reduce circuit fidelity, corrupt gradient estimates, and eventually slow or halt convergence, including through noise-induced barren plateaus~\cite{cerezo2021,wang2021}. Periodic facility-level maintenance, such as nightly recalibration, can restore device quality but this is independent of the state and timing of the running workload. Conversely, runtime calibration can intervene when the workload needs it, but it consumes actual wall-clock time that would otherwise be available for algorithm execution. Existing software-level schedulers, such as Qoncord~\cite{qoncord2024}, while capable of addressing heterogeneous VQA execution, largely treat backend quality as static. Therefore, changes in calibration quality over time remain an architectural blind spot.

Recent developments in control stacks have made this issue particularly important. At the device level, Marciniak \textit{et al.}~\cite{marciniak2026} achieved millisecond-scale autonomous calibration of superconducting qubits using FPGA-based closed-loop control. They demonstrated amplitude tuning in $1\,\mathrm{ms}$, readout optimization in $100\,\mathrm{ms}$, and more than $74{,}000$ recalibrations in six hours. At the system level, tightly coupled HPC--QPU architectures like NVQLink~\cite{NVQLink2025} have reduced quantum--classical round-trip time (RTT) to the microsecond regime. They enable fast interaction between quantum controllers and accelerated classical resources. Therefore, tight-coupling platforms and real-time GPU--QPU interfaces make low-latency intervention technically available. What remains unclear is under what conditions (drift, workload, and calibration-capacity) such intervention can improve end-to-end algorithmic progress.

Our work complements the two levels mentioned above~\cite{marciniak2026,NVQLink2025}. Device-level research shows that fast calibration primitives are achievable, and schedulers like Qoncord~\cite{qoncord2024} primarily address workload placement under the assumption that backend quality remains largely constant. We, however, focus on the intermediate runtime-systems problem: once calibration primitives exist, when should a quantum--classical workflow spend actual wall-clock time to invoke them? Thus, the core question is not whether calibration improves the next circuit, but whether spending time on calibration now can significantly improve the future optimization trajectory, and then justify its immediate cost. A short-sighted runtime scheduler can easily reject effective calibration actions because it only sees the wall-clock loss at the current step, ignoring the ongoing benefit of moving the device to a fresher future trajectory. In runtime calibration policies we therefore propose treating runtime calibration as a state-trajectory intervention, not as a one-off reward.

In an MQSS-style software stack~\cite{mqss2026}, such a runtime calibration policy would sit between the quantum resource manager and compiler infrastructure (QRM\&CI) layer and the QDMI device layer~\cite{qdmi2024}. It uses dynamic device information (such as calibration data, timing constraints, and backend status) exposed through QDMI and FoMaC support libraries~\cite{qdmi2024,FoMac} and returns a runtime decision (such as whether to continue execution, apply a light recovery, or apply a heavier recovery). In this role, runtime calibration becomes a realtime decision visible to the scheduler inside the hybrid loop. This would avoid an external maintenance event or a static backend quality assumption.

In this paper, we formulate the runtime calibration problem as a feedback-control problem under a finite wall-clock budget. We represent calibration quality proxy with an equivalent-age state, model calibration actions as costly state-reset operations, and evaluate policies by the time-integrated optimization gap. Using a finite-horizon rollout model~\cite{bertsekas2020rollout, GARCIA_mpc_survey, rawlings2017book}, we compare runtime calibration with strengthened open-loop baselines, including no calibration and multiple periodic light and heavy schedules. This formulation reveals when trajectory-aware runtime control can outperform static scheduling, and when the overhead of feedback makes it uncompetitive.

Our results clarify the role of latency in the architecture. Lower quantum--classical latency does not automatically guarantee the advantages of runtime calibration. Instead, it extends the effective operational region of feedback-based calibration, making it fast and efficient enough to improve the future trajectory. For isolated single-target recovery, millisecond-scale local control already provides most of the available benefit. The results suggest that the unique advantage of microsecond-scale tight-loop integration should emerge most clearly in capacity-limited regimes where multiple qubits, couplers, gates, or calibration primitives must be processed within the same classical control window.

The main contributions of this paper are as follows:
\begin{itemize}[leftmargin=*]
    \item We introduce a latency-aware equivalent-age model that treats the freshness of calibration quality as a drifting state variable under a fixed wall-clock budget. By formulating runtime calibration as a state-trajectory feedback-control problem, we show that finite-horizon rollout policies can account for state-reset benefits missed by greedy one-step policies.
    \item We compare cloud-like ($25\,\mathrm{ms}$), local-ms ($1\,\mathrm{ms}$), and tight-loop ($4\,\mu\mathrm{s}$) regimes with strengthened open-loop baselines, and show that reduced latency can expand the region of runtime advantage, but does not guarantee it fully. Effective runtime calibration requires timescale matching between drift, calibration overhead, workload sensitivity, and remaining execution time.
    \item Our study of the single target model shows that local-ms can already provide much of available benefit. This motivates capacity-limited multi-target calibration as the next setting, where tight-loop integration would be able to separate more strongly.
\end{itemize}

The remainder of this paper is organized as follows. Section~\ref{sec:model} introduces the latency-aware control model and the equivalent-age state representation. Section~\ref{sec:feedback} formulates runtime calibration as a finite-horizon feedback-control problem and compares it with short-sighted and open-loop baselines. Section~\ref{sec:results} presents numerical results for different latency regimes and workload timescales. Section~\ref{sec:discussion} discusses scalability, capacity limits, and model constraints. Section~\ref{sec:conclusion} summarizes the entire paper.


\section{Model: Calibration as State-Trajectory Feedback Control}
\label{sec:model}

We use a minimal phenomenological model to address the wall-clock trade-offs introduced by runtime calibration. This model is not specific to any particular quantum hardware. It is a regime-level mechanism model that isolates three general characteristics of drifting quantum hardware: calibration quality ages over physical time, recovery actions consume wall-clock time, and the value of a recovery action persists throughout subsequent circuit executions. Instead of separately tracking device parameters such as $T_1$, $T_2$, pulse amplitudes, thresholds, and gate errors~\cite{krantz2019,arute2019quantum,PhysRevApplied.15.034080}, we use an equivalent-age variable to generalize the calibration state.

\subsection{Equivalent age and latency-dependent primitives}
\label{subsec:state_primitives}

Let $a_t \geq 0$ denote the effective calibration age at physical time $t$, defined as the actual wall-clock time elapsed since the processor was last in an ideal calibrated state. The corresponding calibration quality proxy, which we call \emph{freshness}, is a bounded scalar $L_2(a_t)\in(0,1]$, where $L_2=1$ represents an ideal calibration state. We use a Hill-type drift map as an abstraction of time-dependent degradation:
\begin{equation}
    L_2(a_t)
    =
    \frac{1}{1+\left(a_t/\tau_{\mathrm{drift}}\right)^\nu},
    \label{eq:l2}
\end{equation}
where $\tau_{\mathrm{drift}}$ sets the characteristic drift timescale and $\nu$ controls the decay profile.

The inverse map
    $a(L_2)
    =
    \tau_{\mathrm{drift}}
    \left(L_2^{-1}-1\right)^{1/\nu}$
allows calibration recovery to be represented as an equivalent-age reset. After a recovery action, the processor will continue to drift from a younger effective age. We normalize the ideal device-quality factor to $L_1=1$. Since $L_1$ is the same for all policies in a single-backend comparison, it does not affect the relative ranking of the policies, and is omitted below.

At each decision point, the controller can choose a calibration action $p \in \{\varnothing,\mathrm{light},\mathrm{heavy}\}$. The action $\varnothing$ denotes no calibration. A light action represents a fast local correction, such as a single retuning step. A heavy action represents a deeper recovery operation that may require multiple measurement--feedback rounds. Each action $p$ has a target freshness $L_{2,\mathrm{tar}}^{(p)}\leq 1$ and an intrinsic recovery strength $\beta_p\in[0,1]$.

The architectural parameter is the closed-loop round-trip time $\tau_{\mathrm{RTT}}$. We compare three regimes:
cloud-like control with $\tau_{\mathrm{RTT}}\sim \mathcal{O}(10\,\mathrm{ms})$, local-ms control with $\tau_{\mathrm{RTT}}\sim\mathcal{O}(1\,\mathrm{ms})$, and tight-loop control with $\tau_{\mathrm{RTT}}\sim\mathcal{O}(1\,\mu\mathrm{s})$.

A calibration primitive has an intrinsic duration and may require interactive feedback rounds:
\begin{equation}
    T_{\mathrm{cal}}(p,\tau)
    =
    T_{0,p}+N_p\tau,
    \label{eq:tcal}
\end{equation}
where $T_{0,p}$ is the primitive execution time, $N_p$ is the number of feedback rounds, and  $\tau=\tau_{\mathrm{RTT}}$. This form makes heavy multi-round recovery actions more latency-sensitive than light single-round corrections.

\subsection{Time-consistent recovery map}
\label{subsec:recovery}

Calibration consumes physical time, during which the device will continue to drift. If action $p$ starts from age $a_t$, then age will first advance to
    $\tilde a_t
    =
    a_t + T_{\mathrm{cal}}(p,\tau)$.
The pre-recovery freshness is then $L_2(\tilde a_t)$. The action recovers a fraction of the remaining distance to its target freshness:
\begin{equation}
    L_2^{(p)}
    =
    L_2(\tilde a_t)
    +
    \eta_p
    \left[
        L_{2,\mathrm{tar}}^{(p)}
        -
        L_2(\tilde a_t)
    \right]_+ ,
    \label{eq:recovery}
\end{equation}
where $[x]_+=\max(x,0)$ and $\eta_p$ is the realized recovery factor. The recovered freshness is mapped back to an equivalent age
    $a_t'
    =
    a\!\left(L_2^{(p)}\right)$,
so that future drift proceeds from the recovered state. This structure ensures time consistency: calibration costs time, drift accumulates during the action, and the benefit persists into a younger future trajectory.

We distinguish between scheduled recovery and runtime recovery based only on the realized recovery factor. Scheduled maintenance is assumed to execute outside the live feedback deadline, thus achieving the intrinsic recovery strength
    $\eta_p^{\mathrm{sched}}
    =
    \beta_p$.
Runtime recovery is latency-limited. We model the feasibility of a runtime feedback action using the following formula:
\begin{equation}
    L_3(\tau,p)
    =
    \frac{1}{1+\tau/\tau_p},
    \label{eq:l3}
\end{equation}
where $\tau_p$ is the timing tolerance of primitive $p$. The runtime recovery factor is
    $\eta_p^{\mathrm{rt}}
    =
    L_3(\tau,p)\beta_p$.
Therefore, scheduled and runtime calibration use the same recovery map in Eq.~\eqref{eq:recovery}, but their actual efficiencies differ. Importantly, $L_3$ is used only for runtime recovery, not for baseline quality across all policies. This separates the feasibility of feedback actions from the hardware quality and avoids giving low-latency architectures
an artificial global quality boost.

\subsection{Application progress and wall-clock objective}
\label{subsec:objective}
The progress of one algorithm step depends on the instantaneous freshness $L_2$. We define an effective progress factor
\begin{equation}
    Q_{\mathrm{eff}}(L_2,\alpha)
    =
    (1-\alpha)\sqrt{L_2}
    +
    \alpha L_2^\lambda ,
    \label{eq:qeff}
\end{equation}
where $\alpha\in[0,1]$ is used for interpolation between noise-tolerant and noise-sensitive workloads, and $\lambda>1$ controls the nonlinear penalty in the sensitive regime. For small $\alpha$, the concave term indicates applications whose coarse landscape features tolerate moderate degradation. For large $\alpha$, the polynomial term represents gradient or accuracy-sensitive workloads, whose progress is strongly suppressed by drift. In the default evaluation we use $\lambda=2$, so a higher $\alpha$ term imposes a stronger nonlinear penalty when freshness is degraded.

Let $g_t\geq0$ be the normalized optimization gap at algorithm step $t$, with $g_0=1$. The gap approaches a residual floor $g_{\min}>0$ according to
\begin{equation}
    g_{t+1}
    =
    g_{\min}
    +
    (g_t-g_{\min})(1-r_t),
    \label{eq:gap_update}
\end{equation}
where
    $r_t
    =
    \mathrm{clip}
    \left(
        \rho Q_{\mathrm{eff}}(L_{2,t},\alpha),
        0,
        r_{\max}
    \right)$. Here $\rho$ sets the baseline learning rate and $r_{\max}$ bounds the maximum progress in each iteration. The floor $g_{\min}$ represents finite ansatz expressibility, residual noise, and other irreducible constraints of near-term execution.

The simulator evolves over physical time. Calibration time is charged directly to a fixed actual wall-clock budget $T_{\mathrm{budget}}$. Therefore, calibration is only beneficial if the downstream improvement in the gap trajectory compensates for the time delay in algorithm progress.

Policies are evaluated by the mean gap $\overline g$ over physical time,
\begin{equation}
    \overline g
    =
    \frac{1}{T_{\mathrm{budget}}}
    \int_0^{T_{\mathrm{budget}}} g(t)\,dt .
    \label{eq:mean_gap}
\end{equation}
Lower $\overline g$ is better. This objective function naturally penalizes runtime interruptions: during processor calibration, the actual wall-clock time continues to advance and the gap integral continues to accumulate.

For reference, a throughput factor for each iteration can be defined as
    $L_4(p,\tau)
    =
    \frac{T_{\mathrm{alg}}}
    {T_{\mathrm{alg}}+T_{\mathrm{cal}}(p,\tau)}$, where $T_{\mathrm{alg}}$ is the quantum execution time of one iteration. In the fixed-budget simulations, we do not multiply Eq.~\eqref{eq:qeff} by $L_4$, because calibration cost is already included in elapsed physical time. 

The effects of these quantities are intentionally separated. The freshness state $L_2$ determines the progress of the algorithm by $Q_{\mathrm{eff}}$, while $L_3$ only takes effect when the runtime action attempts to reset $L_2$. We use $L_4$ only as a diagnostic for interpreting per-action throughput loss.

\begin{figure*}[t]
    \centering
    \includegraphics[width=0.8\textwidth]{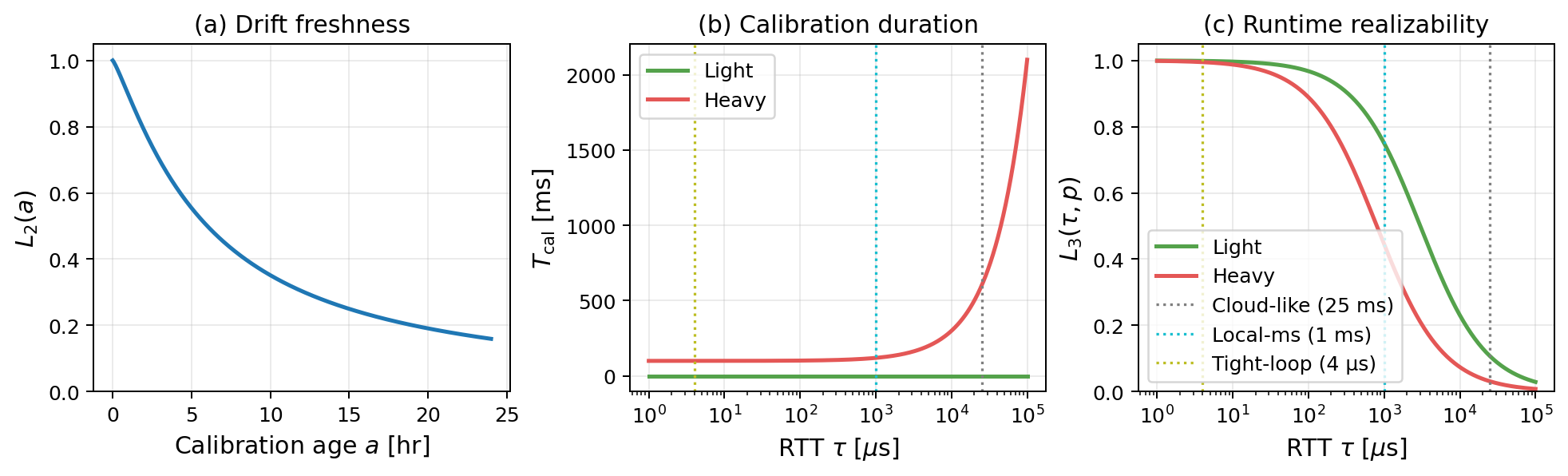}
    \caption{Physical mechanisms underlying the latency-regime
comparison. Longer round-trip time penalizes runtime calibration in
two distinct ways: it increases the duration of interactive
multi-round primitives and reduces the realized recovery efficiency
of feedback actions. The latter penalty is applied only to runtime
recovery, not to offline scheduled calibration or to baseline circuit
quality, which keeps the open-loop comparison fair.
}
    \label{fig:physics_summary}
\end{figure*}

\subsection{State-trajectory value of calibration}
\label{subsec:trajectory_control}

A short-sighted scheduler evaluates calibration only by its immediate effect on the next step. This view is incomplete because calibration is not a transient reward but a state reset. By reducing the equivalent age, a calibration action alters the future drift trajectory and thus affects the execution quality of many subsequent circuits.

The relevant value of calibration is therefore a trajectory-level value. The trajectory value of calibration can be illustrated by the increase in future effective progress factors,
\begin{equation}
    \Delta Q_{\mathrm{traj}}(H)
    =
    \sum_{k=0}^{H}
    \left[
        Q_{\mathrm{eff}}\!\left(L_{2,t+k}^{(p)},\alpha\right)
        -
        Q_{\mathrm{eff}}\!\left(L_{2,t+k},\alpha\right)
    \right].
    \label{eq:traj_value}
\end{equation}
This expression is not the final objective. It summarizes the mechanism:  a state reset raises future $Q_{\mathrm{eff}}$ values. Runtime calibration changes the future hardware-state trajectory, not only the current circuit quality. In the simulations,  this mechanism is evaluated through the finite-horizon gap integral, and final policy performance is measured by the full wall-clock mean gap $\overline g$.

The next section turns this mechanism into a concrete finite-horizon feedback policy and compares it with greedy and open-loop baselines.


\section{Runtime-Control Framework}
\label{sec:feedback}
Sec.~\ref{sec:model} defines how calibration actions change the hardware state. This section connects the state model to the runtime decision problem.

\subsection{Architectural design space}
\label{subsec:model_visual}

We use Figure~\ref{fig:physics_summary} to summarize the mechanism model used throughout the evaluation. The three panels illustrate the sources of runtime calibration value and cost: freshness drift, latency-dependent calibration duration, and runtime feasibility.

As shown in Figure ~\ref{fig:physics_summary}(a), calibration freshness $L_2(a)$ decreases with increasing equivalent calibration age. This represents backend quality as a time-dependent state, not a static processor property. Without intervention, the hardware state ages during the wall-clock execution of the workload.

Figure~\ref{fig:physics_summary}(b) shows how the duration of a calibration primitive depends on the feedback round-trip time $\tau_{\mathrm{RTT}}$. Light primitives are modeled as fast local corrections with weak latency dependence. Heavy primitives require multiple measurement feedback rounds and are therefore more sensitive to latency. As a result, the same physical recovery operation can be affordable in a local-ms or tight-loop regime but too expensive in a cloud-like regime.

Figure~\ref{fig:physics_summary}(c) shows the runtime realizability factor $L_3(\tau,p)$. This factor only affects the recovery efficiency of runtime feedback actions. It is not applied to the baseline circuit quality of all policies. This distinction is crucial: in our model, low latency does not make the processor globally better but rather it makes feedback actions more realizable.

These three mechanisms define the architectural tradeoff. Runtime calibration may improve the future hardware trajectory but also consumes wall-clock time and may be ineffective if the feedback path is too slow. A runtime controller must therefore weigh the downstream trajectory benefit of a calibration action against its immediate cost.

\subsection{Policy classes}
\label{subsec:policies}

We compare five policy classes:
\begin{itemize}[leftmargin=*]
    \item No calibration: the equivalent age grows monotonically throughout the workload.

    \item Periodic heavy calibration: an open-loop schedule applies the heavy primitive at a fixed cadence.

    \item Fixed-cadence light calibration: an open-loop schedule applies the light primitive at a fixed cadence.

    \item Greedy runtime calibration: a feedback policy chooses the action that minimizes the immediate one-step cost. This corresponds to horizon $H=1$.

    \item Trajectory runtime calibration: a finite-horizon rollout policy chooses the action that minimizes the predicted gap trajectory over $H>1$ steps.
\end{itemize}
The first three policies are open-loop baselines. Their actions are fixed by a schedule and are independent of the current freshness state. The last two are feedback policies. Before selecting an action, they observe the current equivalent age, optimization gap, and remaining wall-clock time. 

\subsection{Finite-horizon rollout control}
\label{subsec:rollout_control}

At decision step $t$, the runtime controller observes $x_t = (a_t,g_t,T_{\mathrm{rem}})$, where $a_t$ is the equivalent age, $g_t$ is the current optimization gap, and $T_{\mathrm{rem}}$ is the remaining wall-clock budget. For each candidate first action $p\in\{\varnothing,\mathrm{light},\mathrm{heavy}\} $, the controller applies the physical transition from Sec.~\ref{sec:model} and evaluates a nominal continuation with no further calibration. The rollout sequence is $\mathcal{R}_H(p)
    \equiv
    \left[
    p,
    \underbrace{\varnothing,\ldots,\varnothing}
    _{H-1\ \mathrm{continuation\ steps}}
    \right]$.

The selected action minimizes the predicted finite-horizon gap integral,
\begin{equation}
    p_t^*
    =
    \arg\min_{p\in\{\varnothing,\mathrm{light},\mathrm{heavy}\}}
    \int_{0}^{\min(T_{\mathrm{rem}},T_H)}
    g_{\mathcal{R}_H(p)}(t+s)\,ds .
    \label{eq:rollout-mpc}
\end{equation}
Only the first action $p_t^*$ is executed. Then the state is updated and the remaining time is reduced. The greedy controller is recovered by setting $H=1$. For $H>1$, the controller can assign a calibration action, whose immediate wall-clock cost is high but whose state reset lowers the future gap trajectory.

This rollout controller is simpler than a full recursive finite-horizon optimal-control formulation, such as model predictive control~\cite{rawlings2017book,GARCIA_mpc_survey}. It can be viewed as a receding-horizon rollout with a fixed continuation policy: after the candidate first action, the nominal future policy is no calibration. Thus, the controller does not optimize a complete future sequence of calibration actions. It asks a conservative question: does calibrating now improve the future trajectory relative to doing nothing afterwards? This particular predictive control approach captures the state-investment value of calibration while keeping the online decision cost small enough.

\subsection{Simulation protocol}
\label{sec:protocol}

The simulator evolves in physical time. Each algorithm iteration has a base duration $T_{\mathrm{base}}
    =
    T_{\mathrm{class}} + T_{\mathrm{alg}}$, where $T_{\mathrm{class}}$ is the classical update time and $T_{\mathrm{alg}}$ is the quantum execution time. If a calibration action $p$ is selected, the additional duration $T_{\mathrm{cal}}(p,\tau)$ is also included in the same wall-clock budget. A step is feasible only if it can finish within the remaining budget. If even a no-calibration step cannot be completed, the current gap is integrated over the residual time and the run terminates. 

For a selected calibration action $p\neq\varnothing$, the state transition follows the recovery map in Sec.~\ref{sec:model}: the age first advances during calibration, the freshness is partially restored toward the primitive target, and the recovered freshness
is mapped back to a younger equivalent age. The realized recovery factor is $\eta_p^{\mathrm{sched}}$ for scheduled actions and $\eta_p^{\mathrm{rt}}$ for runtime actions. For $p=\varnothing$, the state simply ages by $T_{\mathrm{base}}$.

After each completed iteration, the optimization gap is updated using Eq.~\eqref{eq:gap_update}. The accumulated gap area is evaluated by a trapezoidal update, $A
    \leftarrow
    A
    +
    \frac{g_k+g_{k+1}}{2}\Delta t_k$, where $\Delta t_k$ is the physical duration of step $k$, including calibration time if calibration is selected. Then the objective is $\overline g = A/T_{\mathrm{budget}}$.

\subsection{Reference baselines and gain metrics}
\label{subsec:gain_metrics}
We compare runtime control with a strengthened open-loop family, not with a single fixed schedule. The open-loop family includes no calibration, several periodic heavy-calibration schedules, and several fixed-cadence light-calibration schedules. This prevents the runtime controller from appearing favorable merely because the baseline schedule was poorly chosen.

For a runtime policy $\pi$, we define the gain:
\begin{align}
    \Delta_{\mathrm{open}}(\pi)
    =
    \min_{\pi'\in\{\mathrm{no\ cal},\mathrm{periodic},\mathrm{fixed}\}}
    \overline g_{\pi'}
    -
    \overline g_{\pi}.
    \label{eq:gain-open}
\end{align}
Positive gain means that runtime control achieves a lower mean gap than the best member of the open-loop reference family. The metric is strict: runtime control must beat the best open-loop choice, including doing nothing.


\section{Numerical Evaluation}
\label{sec:results}

We evaluate runtime calibration under a fixed physical wall-clock budget. All policies use the same state dynamics, recovery map, and objective function defined in Secs.~\ref{sec:model}--\ref{sec:feedback}. The primary metric is the mean optimization gap $\overline g$ from Eq.~\eqref{eq:mean_gap}; lower $\overline g$ indicates better end-to-end progress. Since calibration time is directly charged to the wall-clock budget, a runtime action is effective when its downstream reduction in $g(t)$ compensates for the progress lost during the intervention.

\begin{table}[t]
\caption{Default parameters used in the evaluation. The model is
a regime-level mechanism model; the calibration primitives denote
lightweight and multi-round runtime retuning actions, not full
device tune-up procedures.}
\label{tab:eval_params}
\centering
\renewcommand{\arraystretch}{0.8}
\begin{tabular}{@{}lll@{}}
\toprule
Parameter & Value & Role in the model \\
\midrule
$\tau_{\mathrm{cloud}}$ & 25 ms & cloud-like control latency \\
$\tau_{\mathrm{local}}$ & 1 ms & local control-stack latency \\
$\tau_{\mathrm{tight}}$ & 4 $\mu$s & tight-loop control latency \\
$T_{\mathrm{alg}}$ & 45 ms & quantum time per iteration \\
$T_{\mathrm{class}}$ & scanned; 1 s default & classical time per iteration \\
$\tau_{\mathrm{drift}}$ & 6 h & effective freshness drift scale \\
$T_{0,\mathrm{light}}$ & 1.1 ms & light primitive base time \\
$T_{0,\mathrm{heavy}}$ & 100 ms & heavy primitive base time \\
$N_{\mathrm{heavy}}$ & 20 & rounds in heavy primitive \\
$\beta_{\mathrm{light}},\beta_{\mathrm{heavy}}$ & 0.25, 0.65 & recovery strengths \\
$H$ & 1 or 6 & greedy or rollout horizon \\
$\lambda$ & 2 & freshness-sensitivity exponent\\
\bottomrule
\end{tabular}
\end{table}

Table~\ref{tab:eval_params} lists the default parameters used in the evaluation process. The values define representative latency and calibration timescales, rather than platform-specific fitted values, and the order of magnitude of the input parameters is chosen to be consistent with reported hardware timescales where available. The light primitive is chosen at the millisecond scale to represent fast local retuning, consistent with reported millisecond-scale amplitude tuning. The heavy primitive is set at the $100\,\mathrm{ms}$ scale to represent a deeper multi-round runtime recovery, comparable to reported readout-optimization timescales. The three $\tau_{\mathrm{RTT}}$ values represent cloud-like, local control, and tight-loop coupling regimes. The recovery strengths are phenomenological and are varied in the robustness checks. Unless stated otherwise, we use $T_{\mathrm{class}}=1\,\mathrm{s}$ and compare greedy runtime control with $H=1$ against trajectory runtime control with $H=6$. 

\begin{figure*}[t]
    \centering
    \includegraphics[width=0.8\textwidth]{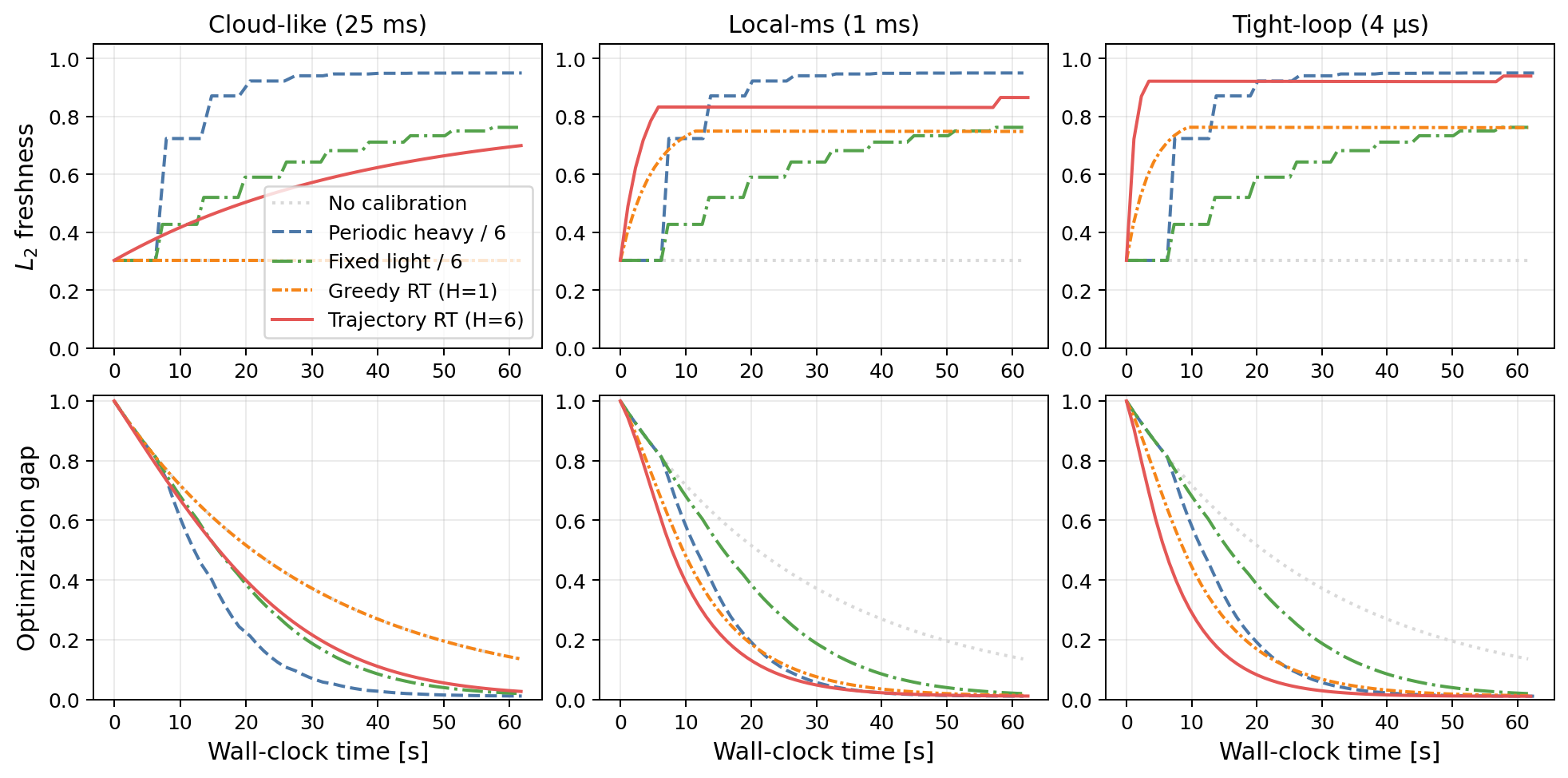}
    \caption{Representative time-domain trajectories at $a_0=12$ h, $\alpha=0.7$, and $T_{\mathrm{class}}=1$ s. Top: $L_2$ freshness; Bottom: optimization gap $g(t)$. Calibration acts as a discontinuous state reset. Runtime calibration is useful only when the state reset lowers the downstream gap trajectory enough to repay its wall-clock cost.
}
    \label{fig:trajectories}
\end{figure*}

The reference set contains no calibration, periodic heavy calibration with periods $\{3,6,12\}$, and fixed-cadence light calibration with periods $\{3,6,12\}$. The scheduled policies are specified in nominal no-calibration iteration periods, but are triggered on the wall-clock timeline. Thus, a slower policy cannot hide calibration overhead by simply counting completed algorithmic iterations. We emphasize that the qualitative comparisons below should be understood as regime-level design studies, not hardware-fitted predictions.

\subsection{Trajectory-level effect of calibration}
\label{subsec:trajectory_traces}

Calibration is not a static backend property; it is a temporal state that decays during execution. Figure~\ref{fig:trajectories} illustrates the trade-off between calibration overhead and performance gain under a fixed wall-clock budget.  A runtime intervention resets the freshness state $L_2$ only after its calibration time has been paid. Its value is therefore measured by the resulting reduction in the area under the $g(t)$ curve, not by the immediate post-action freshness. This chosen point $(a_0=12\mathrm{h},\alpha=0.7$) is used only to make the trajectory mechanism visible, while the full parameter-space region is evaluated separately in Fig.~\ref{fig:gain_maps}.

The representative trajectories show that runtime calibration is architecture-dependent. In the cloud-like regime, a runtime reset of $L_2$ is slow and poorly realized. The trajectory-aware policy can improve freshness, but the time cost is high enough that a strong open-loop schedule remains better. Reducing the round-trip time changes this tradeoff. In the local-ms and tight-loop regimes, runtime feedback becomes fast and feasible enough for a state reset to pay back its wall-clock cost. The greedy controller remains conservative because it evaluates mainly the one-step penalty of calibration. The finite-horizon rollout controller also values the future iterations executed on the fresher state, and therefore accepts calibration when the downstream gap reduction is large enough.

This example also separates the number of calibration actions from their value. Tight-loop feedback need not perform more calibrations; its advantage lies in performing fewer but more effective calibrations. This distinction will be quantified in Sec.~\ref{subsec:action_mechanism} below.

\subsection{Runtime viability against open-loop baselines}
\label{subsec:gain_maps}
\begin{figure*}[t]
    \centering
    \includegraphics[width=0.8\textwidth]{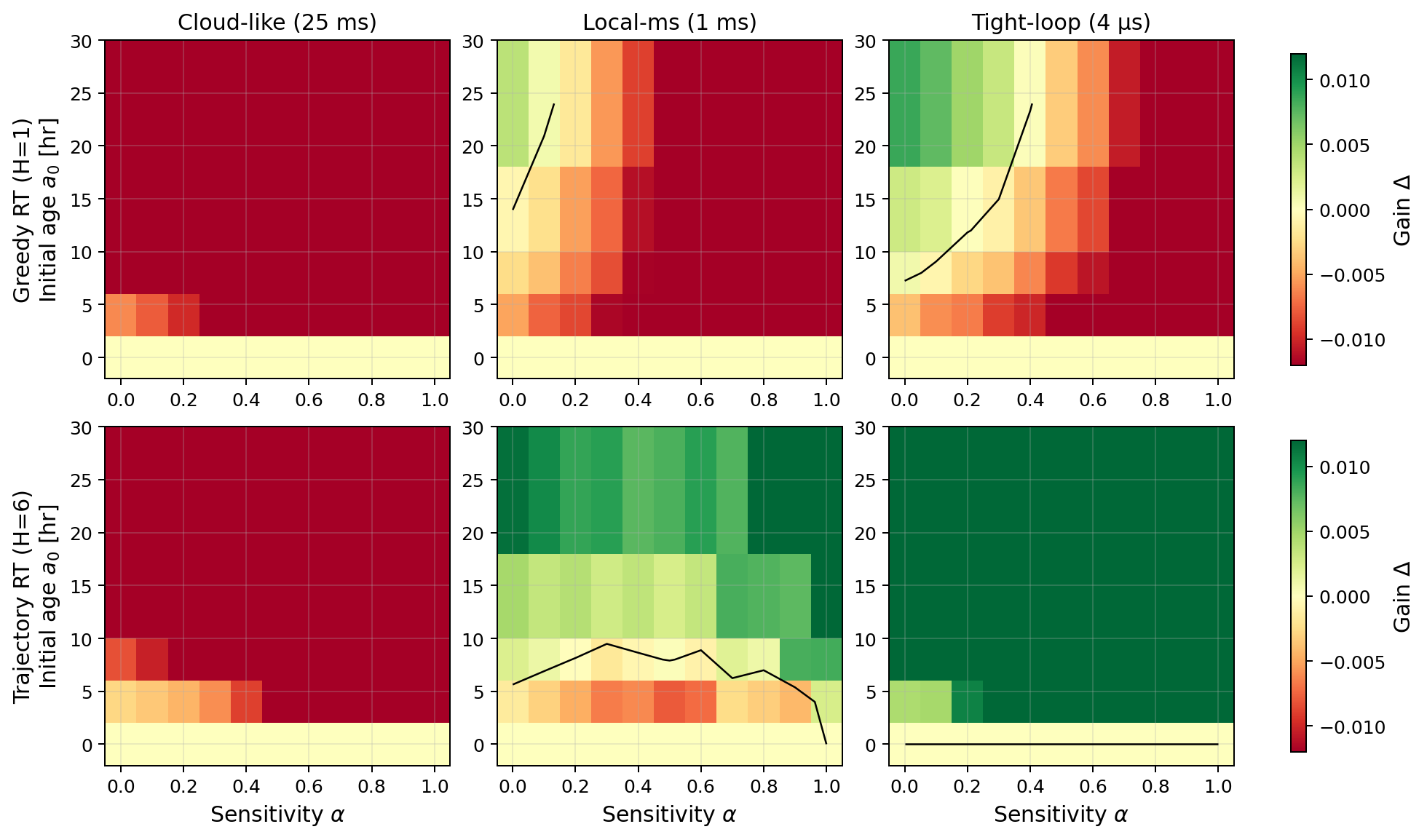}
    \caption{Runtime gain $\Delta_{\mathrm{open}}$ relative to the strengthened open-loop reference family. Top: greedy ($H=1$); Bottom: rollout ($H=6$). Columns correspond to the cloud-like, local-ms, and tight-loop latency regimes. The axes scan algorithmic sensitivity $\alpha$ and initial calibration age $a_0$, respectively. Red regions indicate that runtime control loses to the best open-loop policy; green regions indicate positive runtime gain. The black contour marks $\Delta_{\mathrm{open}}=0$ where visible. Local-ms feedback opens a finite positive-gain region, while tight-loop feedback gives the broadest operating margin in this scalar model.
}
    \label{fig:gain_maps}
\end{figure*}

We evaluate the robustness of our trajectory-aware control across the parameter space. Figure~\ref{fig:gain_maps} illustrates the impact of algorithmic sensitivity $\alpha$ and initial calibration age $a_0$ on $\Delta_{\mathrm{open}}$ under greedy policies ($H=1$) and rollout policies ($H=6$). The columns compare the three round-trip-time regimes. Positive regions indicate that runtime control achieves a lower mean gap than the best member of the strengthened open-loop reference family. The results clearly show an ordering of the latency regimes: cloud-like runtime feedback is generally uncompetitive, local-ms feedback opens a finite positive-gain region, and tight-loop feedback gives the broadest region.

Runtime-calibration value is strongly architecture-dependent and constrained by the interplay between control latency and workload sensitivity. In the cloud-like regime, both greedy and trajectory-aware feedback are almost entirely negative. This negative result means high-latency runtime feedback is often not worth executing when compared to a strong open-loop reference.

The local-ms regime opens up a nontrivial positive-gain region, especially for freshness-sensitive workloads and aging devices ($a_0>0$). Tight-loop feedback expands this region further and gives the largest positive margins in the scalar model. Lower round-trip time increases the margin in which runtime recovery is sufficiently fast and achievable to improve the future trajectory. When the device is already fresh, little recoverable headroom is available and no runtime policy gains much.

The row-wise comparison provides the main control result. Greedy control is generally conservative because it sees calibration mainly as an immediate throughput loss. The rollout controller, by contrast, values the persistent state reset and is therefore more robust in the local-ms and tight-loop regimes.

\begin{figure}[!t]
    \centering
    \includegraphics[width=0.8\columnwidth]{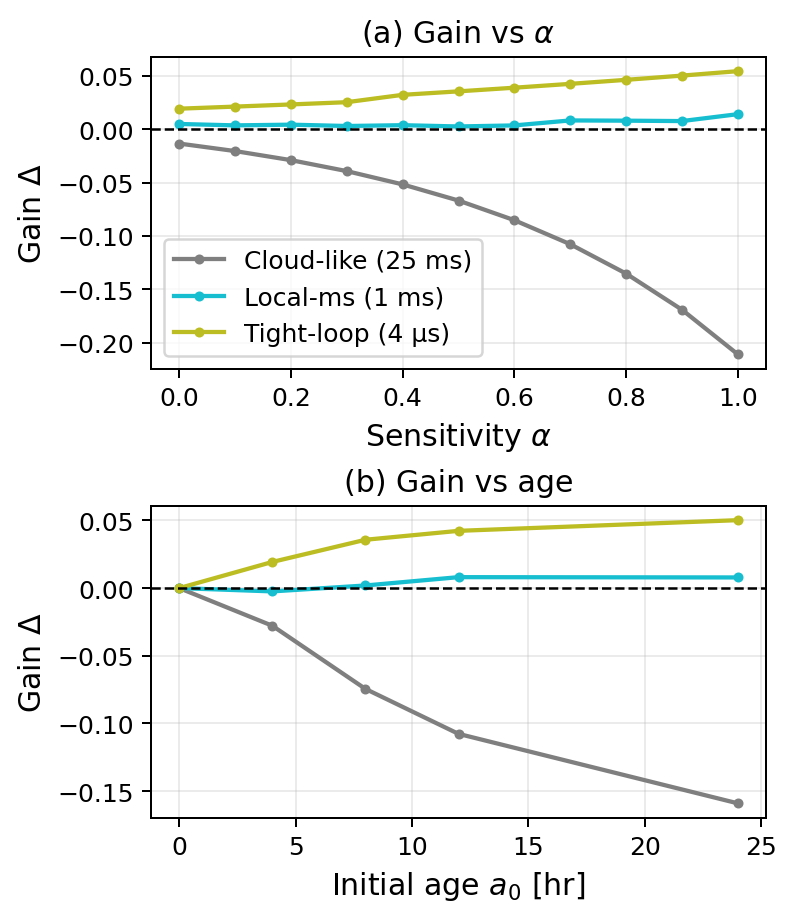}
    \caption{One-dimensional slices of trajectory-runtime gain. (a) Gain versus algorithmic sensitivity $\alpha$ at fixed initial age $a_0=12\,\mathrm{h}$. (b) Gain versus initial calibration age $a_0$ at fixed sensitivity $\alpha=0.7$. Tight-loop feedback gives the largest positive margin, while cloud-like feedback becomes increasingly unfavorable as sensitivity or recoverable age increases.
}
    \label{fig:regime_curves}
\end{figure}

Figure~\ref{fig:regime_curves} clarifies these trends through fixed one-dimensional slices. Panel (a) varies $\alpha$ at
$a_0=12\,\mathrm{h}$, while panel (b) varies $a_0$ at $\alpha=0.7$. In the low-latency regimes, the gain increases as the workload becomes more freshness-sensitive and as the device provides more recoverable headroom. The effect is strongest for tight-loop feedback, while local-ms feedback remains close to zero or weakly positive. Cloud-like feedback follows the opposite trend: it becomes more unfavorable as sensitivity or initial age increases, because the runtime recovery cost cannot be repaid. When the device is already fresh, little gain is available for any policy.

\subsection{Action diagnostics}
\label{subsec:action_mechanism}

\begin{figure*}[t]
    \centering
    \includegraphics[width=0.8\textwidth]{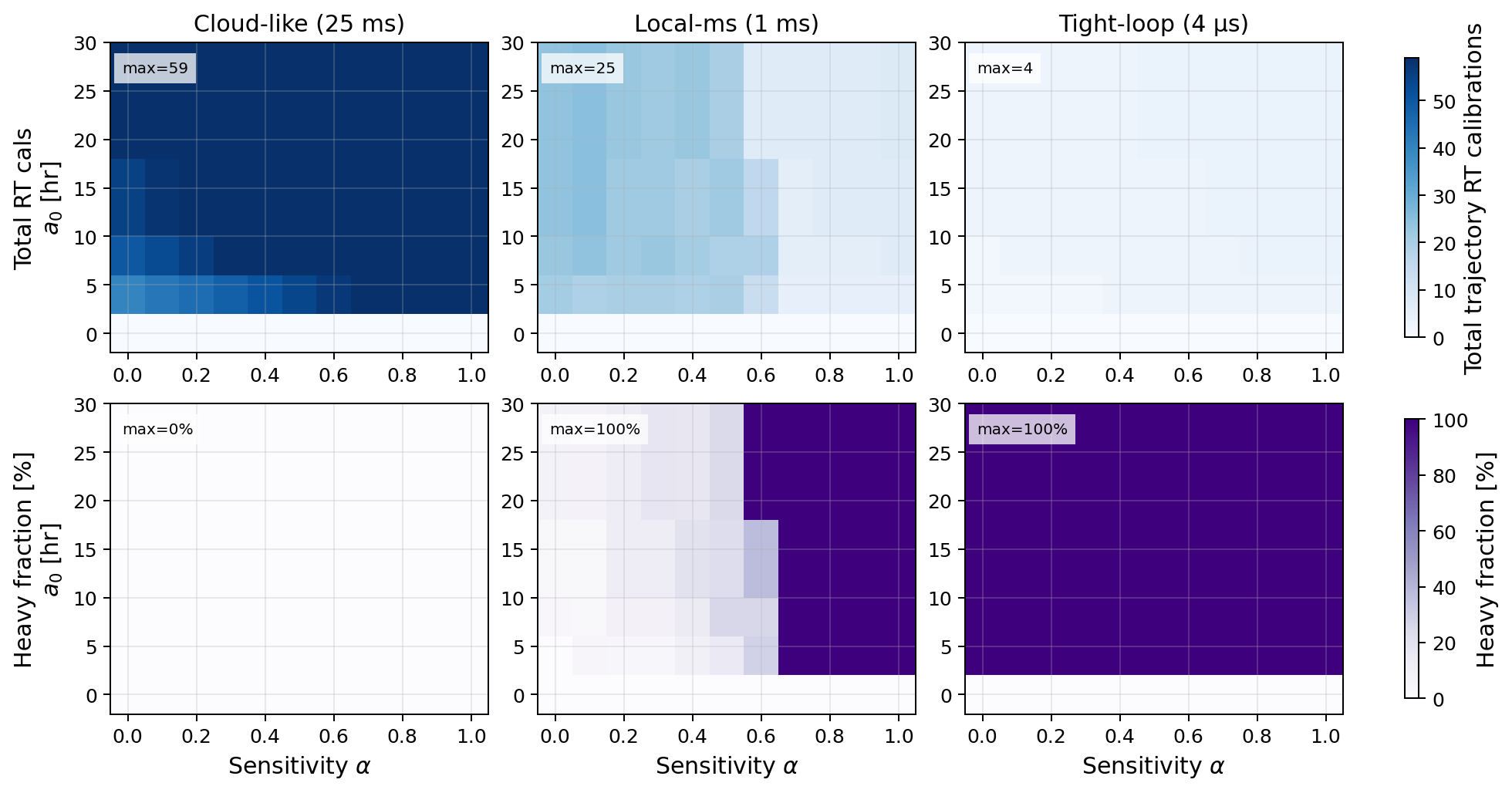}
    \caption{Action diagnostics for the trajectory runtime controller ($H=6$). Top: Total runtime calibration actions executed within the fixed budget. Bottom: The fraction of those actions classified as heavy primitives.  Cloud-like feedback can trigger many calibrations, but they are almost entirely light actions. Tight-loop feedback selects far fewer actions, but they are predominantly heavy state resets. Runtime-control quality is therefore determined by intervention value, not by invocation count alone.
}
 \label{fig:action_mechanism}
\end{figure*}

To understand the mechanics driving the trajectory controller ($H=6$), Figure~\ref{fig:action_mechanism} analyzes the policy behavior by decoupling intervention frequency from intervention depth. We report the total number of runtime actions executed within the fixed wall-clock budget (top row) and the fraction of those actions that use heavy, multi-round primitives (bottom row). 

The failure of the cloud-like regime is not due to controller inactivity. On the contrary, the high-latency policy triggers a large number of calibrations, but these operations are exclusively light. With a $25\,\mathrm{ms}$ round-trip time, heavy recovery is too costly and too weakly realized to repay its state reset value. The controller is therefore left with frequent shallow interventions that do not beat the strengthened open-loop reference.

Tight-loop integration exhibits the opposite pattern. With microsecond-scale round-trip time, heavy state resets become highly feasible and affordable. The controller consequently selects far fewer actions, yet relies almost entirely on heavy primitives. The local-ms architecture represents the intermediate transition: heavy actions appear only in the parts of the grid where their state-reset value compensates their time cost. Thus, runtime value is governed by the depth and realizability of the selected intervention, not by the raw number of controller invocations.

\subsection{Classical-loop timescale sensitivity}
\label{subsec:classical_scan}

\begin{figure}[!t]
    \centering
    \includegraphics[width=0.8\columnwidth]{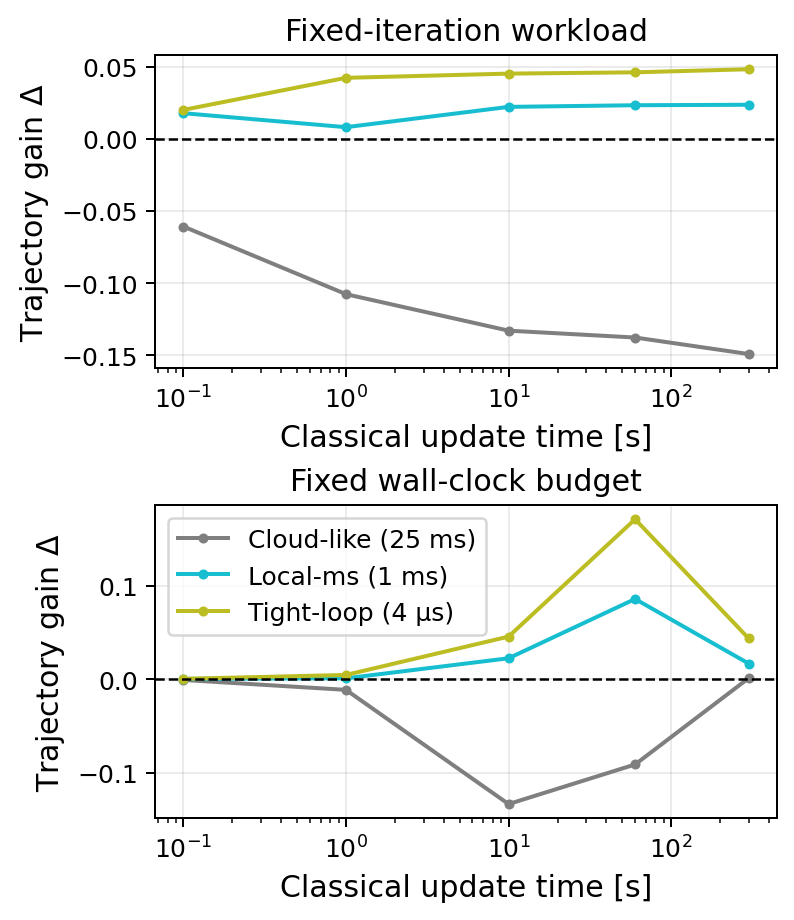}
    \caption{Classical-loop timescale sensitivity for trajectory runtime control. Top: fixed-iteration workload, where the wall-clock budget grows with $T_{\mathrm{class}}$. Bottom: fixed wall-clock budget, with $T_{\mathrm{budget}}=600\,\mathrm{s}$. In the fixed-wall-clock scan, runtime gain is non-monotonic because calibration must be spread over the remaining future iterations. Positive gain is controlled by the matching between calibration overhead, device drift, and algorithm-loop cadence.
}
    \label{fig:classical_time_scans}
\end{figure}

The value of a runtime state reset depends on the cadence of the surrounding classical loop, $T_{\mathrm{class}}$. Figure~\ref{fig:classical_time_scans} isolates this dependence using two workload definitions. In the fixed-iteration scan, the wall-clock budget grows with $T_{\mathrm{class}}$, thus preserving the nominal number of no-calibration iterations. In the fixed-wall-clock scan, the budget is fixed, so slower classical updates reduce the number of future iterations available to benefit from calibration.

In the fixed-iteration scan, local-ms and tight-loop feedback remain positive across the scanned classical-loop times, while cloud-like feedback remains negative against the strengthened open-loop reference. This shows that low-latency feedback can retain value when the workload length grows with the classical update time.

The fixed-wall-clock scan is more restrictive and introduces a significant window effect. The gain becomes non-monotonic here. If the classical loop is very fast, calibration can be expensive relative to a single iteration. If the classical loop is very slow, the remaining iterations are too few to offset the intervention cost. Therefore, runtime calibration is most valuable in the intermediate regime where calibration overhead, drift, and algorithm-loop cadence are comparable. 

In this scan, cloud-like feedback remains mostly noncompetitive, whereas local-ms and tight-loop feedback exhibit a clear positive-gain window. Thus, lower round-trip time expands the opportunity for feedback gain, but does not determine it alone. Runtime calibration also requires recoverable drift, freshness-sensitive workload dynamics, and enough remaining trajectory length to repay the calibration cost.

\section{Discussion}
\label{sec:discussion}

The application objective in our model is relatively abstract. We use the time-integrated optimization gap as a compact metric. This isolates the wall-clock tradeoff, but cannot directly predict specific quantum-chemistry energy errors or application-level fidelity. Accordingly, the numerical gain values should be interpreted as normalized differences in the model objective, not as direct application-error estimates.

We also performed robustness scans over the rollout horizon, drift timescale, recovery strengths, heavy-primitive feedback rounds, and the functional form of $L_3(\tau,p)$ (rational, exponential, and linear-cutoff functions). The scans maintain the qualitative regime ordering. Cloud-like feedback remains mostly uncompetitive, local-ms feedback is viable but parameter-sensitive, and tight-loop feedback provides the broadest operating margin in the scalar model. We also repeated the main sweep with $\lambda=0.6$ instead of the default $\lambda=2$ in $Q_{\mathrm{eff}}$. This changes the quantitative gain boundaries but preserves the qualitative latency-regime ordering. The value of runtime calibration depends on the workload timescale. If the classical loop is very fast, calibration can be too expensive relative to one algorithm iteration. If the classical loop is very slow, too few future iterations remain to compensate for the calibration investment. 

The present single-state model is conservative. For a single effective calibration target, the difference between local-ms and tight-loop control may be modest. If the classical algorithm loop is much longer than the calibration action, and if a light recovery primitive is sufficient, a millisecond-scale local controller may already provide most of the available runtime benefit. This is consistent with our results here. Once the relevant state can be refreshed within the useful trajectory window, further latency reduction gives diminishing returns. However, real devices do not have only a single calibration target. Runtime calibration demand is distributed across qubits, couplers, resonators, primitive-specific dependency constraints, and full calibration dependency graph~\cite{krantz2019,kelly2018,arute2019quantum,PhysRevApplied.15.034080}. Under such capacity pressure, the relevant question is not only whether one correction is fast enough, but also how many active targets can be serviced within the same control window. In this context, the bottleneck is no longer the recovery of one state variable, but the service rate of multiple coupled calibration targets. A relevant system metric would be to measure the calibration service capacity, in which one measure the number of active calibration targets (such as state estimations and recovery operations) that can be completed within a given control window. Then the system performance can be determined by the balance between calibration demand and latency-constrained process throughput. As system size and workload complexity increase, calibration demand scales with the number of active components need to be processed concurrently. In this situation, tight-loop architectures can then provide more service capacity for multi-target recovery before the wall-clock budget runs out. 

This study is not limited to superconducting processors, but the relevant timing domain depends on the platform. NVQLink distinguishes between latency-sensitive systems and latency-insensitive systems. Latency-sensitive systems require tightly scheduled workflows, while latency-insensitive systems allow for more real-time host coordination and interactive execution~\cite{NVQLink2025}. This distinction is crucial for neutral-atom and trapped-ion platforms, because their dominant state variables and recovery primitives differ from superconducting devices~\cite{henriet2020neutral,bruzewicz2019trapped}. For example, freshness may be affected by atom loss, rearrangement quality, or laser stability, and some recovery actions such as rearrangement, recooling, or reloading can be much slower than the speed of the superconducting feedback loop. In such cases, microsecond-scale round-trip time may not dominate a single recovery action.

\section{Conclusion}
\label{sec:conclusion}
Our core conclusion is that runtime calibration is only effective if it can significantly alter the future hardware state trajectory to offset its time-consuming costs. In this view, calibration is not merely a reward for a single-step operation, but a state reset operation. It alters the quality of future circuit execution by reducing the device's equivalent age. Our results also indicate that runtime calibration is is highly dependent on architecture and workload. Cloud-like feedback is generally uncompetitive, whereas local-ms and tight-loop regimes can produce positive runtime gain when the device has recoverable drift, the application is sensitive to freshness, and future execution remains enough to offset the intervention.

Reducing quantum-classical latency does not automatically improve end-to-end performance. It expands the regime in which runtime recovery is fast and realizable enough to improve the future hardware-state trajectory. For isolated single-target workloads, local-ms feedback may already provide much of this benefit. The distinctive advantage of tight-loop integration is expected to emerge most clearly under capacity pressure, where multiple calibration targets must be processed within the same classical control window.

\section*{Acknowledgment}
The author thanks Martin Schulz for helpful feedback on the related work and software-stack positioning of this work, and colleagues and collaborators at LRZ, MQV, and IQM for valuable discussions on HPC--QC runtime systems and quantum device calibration processes. 

This work was supported by the German Federal Ministry of Research, Technology and Space (BMFTR) under grant 13N16690 (Euro-Q-Exa); by the European Union under grant 101194491 (QEX); and by the Bavarian State Ministry of Science and the Arts (StMWK) through Munich Quantum Valley (MQV), including Q-DESSI.

The author used ChatGPT (OpenAI) to assist with language editing and clarity improvements. The author reviewed, revised, and validated all content and takes full responsibility for the work.

\bibliographystyle{IEEEtran}

\bibliography{runtime}

\end{document}